\documentstyle[prb,aps,epsfig,preprint]{revtex}
\begin{document}
\title{The band structure of BeTe - a combined experimental and theoretical
study}
\author{M.~Nagelstra\ss{}er, H.~Dr\"oge, H.-P.~Steinr\"uck}
\address{Experimentelle Physik II, Universit\"at W\"urzburg, 
D-97074 W\"urzburg, Germany}
\author{F.~Fischer, T.~Litz, A.~Waag, G.~Landwehr}
\address{Experimentelle Physik III, Universit\"at W\"urzburg, 
D-97074 W\"urzburg, Germany}
\author{A.~Fleszar and W.~Hanke}
\address{Theoretische Physik I, Universit\"at W\"urzburg, 
D-97074 W\"urzburg, Germany}
\maketitle
\begin{abstract}
Using angle-resolved synchrotron-radiation photoemission spectroscopy we have 
determined the dispersion of the valence bands of BeTe(100) along $\Gamma X$, 
i.e. the [100] direction. The measurements are analyzed with the aid 
of a first-principles calculation of the BeTe bulk band structure as well as 
of the photoemission peaks as given by the momentum conserving bulk 
transitions. Taking the calculated unoccupied bands as final states of the 
photoemission process, we obtain an excellent agreement between experimental 
and calculated spectra and a clear interpretation of almost all measured bands.
In contrast, the free electron approximation for the final states fails to 
describe the BeTe bulk band structure along $\Gamma X$ properly. 
\end{abstract}
\vspace*{2cm}
PACS: 71.20.Nr, 71.55.Gs, 79.60.-i
\newpage
\section{Introduction}

The current activities concerning electro-optical devices in the blue-green 
spectral region have led to a significant interest in the electronic and 
geometric properties of II-VI semiconductors, in particular of epitaxial 
II-VI heterostructures. These materials can be grown on various substrates 
in a well defined fashion by molecular beam epitaxy (MBE)\cite{c1}. Recently, 
the new II-VI compound material BeTe has received considerable attention and 
it has been demonstrated that epitaxial layers of high structural 
quality can be produced by MBE. BeTe crystallizes in the zincblende 
structure and has a number of technologically interesting properties which 
make it attractive for the use in opto-electronic devices:
Its lattice constant of 5.627 $\AA$  is very close to that of 
GaAs and ZnSe (mismatch smaller than 0.7 \%)\cite{c2}. Furthermore, BeTe has 
a large bond energy and a much higher degree of covalency than other wide-gap 
II-VI semiconductors such as CdTe or ZnSe which results in an 
increased hardness and stability\cite{c3}.\\

For the development and optimization of a material in technological 
applications, i.e. for a controlled 
band engineering in superlattices and quantum well structures, its 
electronic structure and in particular its k-resolved bulk and surface 
band structure have to be known in detail. The most powerful experimental 
method to determine the dispersion $E(\vec{k})$ of the various occupied bands 
along a particular direction in k-space and to obtain 
absolute critical point energies is angle-resolved photoelectron 
spectroscopy utilizing linearly polarized synchrotron 
radiation\cite{c4,c5,c6,c7,c8}.\\

In contrast to other II-VI semiconductors, BeTe has been rarely studied 
experimentally up to now and very little is known about its various 
physical properties, in particular its electronic structure. Also, theory has 
not paid much attention to BeTe and only few band structure calculations 
exist\cite{c9}. In this paper we present the first angle-resolved photoemission
study to investigate the k-resolved band structure of BeTe along $\Gamma X$, 
i.e. the (100) direction. 
The measurements were accompanied by and analyzed with state-of-the-art 
first-principles calculations of the BeTe bulk band structure and theoretical 
photoemission spectra. This combined approach turned out to be very successful.
We can identify the electronic transitions which give rise to particular 
photoemission peaks and are able to provide both experimentally and 
theoretically the structure of occupied states in BeTe. Both approaches are in 
good agreement with each other. Furthermore, we find that the assumption of 
free-electron final states for the photoemission process breaks down in the 
case of BeTe. Our results suggest that a much better understanding of the 
origin of the photoemission peaks can be obtained with the application of 
first-principles calculation not only as a theoretical result to compare with, 
but rather as a tool to analyze the experimental measurements.\\
  
\section{Experimental}

The angle-resolved photoemission experiments were performed at the German 
synchrotron radiation facility BESSY in Berlin using linearly polarized 
synchrotron radiation in the energy range from 10 to 19 eV using an 
UHV system of the Technical University Munich that operates at a base 
pressure of better than $1\times10^{-10}$ mbar. The electron analyzer is a 
homebuilt angle-multichannel instrument that allows for a simultaneous 
detection of 
electrons emitted in the polar angle range from 0 to $90^\circ$, for a fixed 
azimuth\cite{c10}; its polar angle resolution is $2^\circ$ with an azimuthal 
acceptance of $3^\circ$. The combined energy resolution of monochromator and 
electron energy analyzer, as determined from the width of the Fermi edge of 
an gold sample, 
varied from 0.15 to 0.25 eV with increasing photon energy. All spectra 
shown below have been measured at normal incidence of the 
incoming radiation. Electron binding energies are referred to the valence 
band maximum (VBM) of BeTe.\\

The epitaxial BeTe(100) layer were produced by molecular beam epitaxy (MBE) 
under ultrahigh vacuum (UHV) conditions in a Riber 2300 four-chamber system 
at the university of W\"urzburg. The growth chambers are equipped with 
reflection high-energy electron-diffraction (RHEED) instrumentation for in 
situ characterization. The GaAs(100) substrate ($8\times8 mm^2$) was mounted on
a special inlay that can be attached onto the 2" sample holder of the 
MBE-system or the manipulator of the UHV electron spectrometer. Its temperature 
was measured by a thermocouple in contact with the molybdenum block holding 
the inlays. Epitaxial GaAs buffers ($500\ nm$) have been used to improve 
the quality of the n-doped GaAs(100) substrate surface. A BeTe 
layer of 100 nm thickness was grown at a substrate temperature of $300^\circ$ 
C in a Te-rich regime as confirmed by the $(2\times1)$ reconstruction observed 
by RHEED. The strain of the layer is unknown. Other BeTe layers of this
thickness have been found to be partly relaxed. For the present study a fully
relaxed layer has been assumed.  One should note, that even for a not fully 
relaxed layer no significant changes in the electronic structure (within the 
accuracy of our study) are expected due to the small lattice mismatch to GaAs 
of $0.7$ \% .\\

After the growth process the samples were transported from the MBE system 
in W\"urzburg to the electron spectrometer in Berlin with a specially 
designed UHV transfer box that can carry up to 6 different 
samples. The pressure in the transfer box that is pumped by an ion getter pump 
is better than $3\times10^{-9}$ mbar. After the transfer from the MBE system 
to the electron spectrometer at BESSY, the BeTe(100) surface showed a sharp 
$(2\times1)$ LEED pattern and there was no 
indication of contamination, as judged by Auger-electron spectroscopy.\\
 
\section{Calculations}

The experimental studies were accompanied by a first-principles calculation 
of the BeTe band structure within the framework of the density-functional 
formalism in the local-density approximation (LDA)\cite{HKS}.
The electron-ion interaction was modelled by norm-conserving, Be$^{2+}$ and
Te$^{6+}$ pseudopotentials generated according to the scheme of 
Hamann\cite{c11} taking into account the partial core-charge correction 
in the exchange-correlation energy functional\cite{c12}.
Because the spin-orbit interaction is known to be very important in tellurides,
in particular for experiments probing details of the band structure, we have
explicitly included the spin-orbit interaction in
the hamiltonian after a selfconsistent solution of the Kohn-Sham equations has 
been obtained.
A plane-wave basis set was employed with the cut-off of 14 Rydbergs.\\

A rather common practice for the assignment of a wave vector to a measured
photoemission peak is to use free-electron, parabolic bands along 
the surface normal as final states for the photoemission excitation process.
In a more sophisticated approach the calculated bands (possibly within some
ab initio scheme) are assumed to be the final states. However, even such 
a procedure does not unambiguously explain the origin of measured peaks,
because in the case of a more complicated band structure of the unoccupied 
states - and BeTe is such a case - the considerations based on the 
energy-conservation principle alone, without
taking into account the transition matrix elements and specific photoemission
selection rules are not sufficient. Therefore, in order to trace back the
origin of various measured structures, we have calculated the 
electron-hole excitation rate for BeTe bulk, which gives an essential
contribution to the emitted-photocurrent intensity.
We have calculated the following quantity\cite{c5}:\\

\begin{equation}
I(E,h\nu) \sim \sqrt{E_{kin}} \sum_{i,f} \lambda_f \int dk_{\perp} 
| \langle \psi_{\vec{k}i} | \vec{A} \cdot \vec{p} | \psi_{\vec{k}f}\rangle | ^2 
\delta(E_f - E_i - h\nu) \delta(E-E_i), 
\end{equation}\\

where $\psi_{\vec{k}i}$, $E_i$, and $\psi_{\vec{k}f}$, $E_f$ are the Kohn-Sham
wave functions and energies of initial and final states in the bulk.
$I(E,h\nu)$ is calculated as a function of the binding energy of 
occupied states $E$ and incident photon energy $h\nu$;
$\vec{k}=(\vec{k}_{\parallel}=0,k_{\perp})$ for normal emission.
$\lambda_f$ is the inelastic mean free path of final states and $E_{kin}$
is the kinetic energy of detected photoelectrons, which is in our calculation
taken as the excess energy of the final states over the vacuum-level energy.\\

It should be noted that although the above expression complies with the
energy conservation rule and takes into account the transition-matrix elements
as well, it includes only rather approximatively 
the presence of the surface of the solid and the many-body effects
in the initial and final states. In particular, the surface effects are
very important. First, the surface gives rise to
the presence of surface states and resonances, which usually show
up in the photoemission spectra as $h\nu$-independent peaks. Such states
are neglected in our first-principles calculation and we will therefore
focus on the identification of bulk peaks in the experimental spectra.
Second, final states of photoelectrons are scattering states with a proper
free-electron-like character far outside the surface\cite{c5}. For such states
the $k_{\perp}$-conservation rule does not apply and non-vertical transitions
are present. Moreover, these states can have their energies within the
excitation gaps in the solids, leading to unexpected and dispersionless
photoemission peaks\cite{c15}. Assuming a common, three-step
model of the photoemission experiment, Eq.(1), in which 
the bulk states are taken to be the final states, can be viewed as 
an approximation in which the transition probability through the surface
barrier is neglected. In order to comply - at least partially - with the
surface barrier transition probability and the specific photoemission
selection rules, we have allowed only for those final states in Eq.~(1),
which are symmetric under all $C_{4v}$ group-symmetry operations, i.e. have
no nodes along the normal to the surface symmetry axis\cite{c16}.\\

As far as the many-body effects are concerned, they show up in several ways 
as well. First, they lead to energy shifts of occupied and unoccupied bands.
This effect is to a large extent included in our calculation via the empirical 
shift of conduction bands. A comparison with a calculation done within
the GW approximation\cite{HL} shows that
a rigid shift of bands is a good approximation in the excitation-energy
range relevant for our measurements\cite{c17}. Second, unlike for the 
independent-particle theory, which has been applied in Eq.(1) above, the 
binding- and excitation energies of quasiparticles within a dynamic many-body 
theory are not sharp. The quasiparticle states are described by spectral 
functions which - although possessing distinct peaks - are continuous 
functions of energy. Electrons and holes have finite lifetimes, which lead
to an important relaxation of the energy- and momentum conservation rule.
We are including this effect via replacing the delta functions in Eq.(1) by 
the Lorentzian functions of a width given by the imaginary part of the 
self-energy, which has been calculated within a GW approximation\cite{c17}.
Finally, the finite mean free path of electrons, which appears in Eq.(1), 
originates from the imaginary-part of  self-energy and the velocity of excited 
electrons\cite{c17}.\\

A proper theory of the photoemission should be ideally a {\it one-step} theory,
in which all the surface and many-body effects would be fully included.
Our aim is to obtain as good as possible an understanding of the origin
of the measured photoemission peaks on the basis of a relatively simple
bulk calculation, with an approximative account of surface and many-body
effects. The success of our approach will be discussed in the next Section.\\

\section{Results and Discussion}

The calculated band structure of BeTe with the experimental lattice constant of 
5.627 $\AA$  is presented in Fig.~1. Our calculation shows that BeTe is an 
indirect semiconductor, with the minimum of the conduction band at the 
$X$-point. The LDA absolute gap is 1.62 eV, which is about 1.1 eV smaller than 
the experimentally reported values of 2.7 eV\cite{c18} and 2.8 eV\cite{c1}. 
This is a typical error due to the Kohn-Sham formalism and the 
LDA approximation. A GW-calculation, in which dynamical, many-body interactions
are to a big extent accounted for, gives a value of 2.6 eV for the absolute gap
in BeTe\cite{c17}. Considering the LDA error of 1.1 eV, a value of 4.45 eV
is obtained for the direct gap at the $\Gamma$-point (see also Fig.~4 below),
which is in very good agreement with the experimental value of $4.53 \pm 0.1$ 
eV as determined by ellipsometry\cite{Wagner}.
The calculated spin-orbit splitting at the valence-band 
maximum at the  $\Gamma$ point is 0.96 eV. The calculated effective masses
in the neighborhood of the $\Gamma$-point are $0.34$, $0.23$ and $0.40$ in
units of the free-electron mass $m_{el}$ for the heavy holes, light holes and 
the spin-split-off band, respectively.\\

For the experimental determination of the bulk band structure of BeTe along the
$\Gamma X$ direction, we have measured photoemission 
spectra for the (100) surface at normal emission for photon energies between 
10 and 19 eV. These spectra are depicted in Fig.~2 with the binding energy 
scale referenced to the valence band maximum (VBM), which is 
determined by the peak position of the feature with the lowest binding energy. 
Overall the spectra show a number of well defined maxima that disperse with increasing photon energy, i.e. kinetic energy of the outgoing photoelectron. For four peaks the positions
of the peak maxima are indicated with different symbols. These 
peaks are assigned to direct transitions from the occupied to the unoccupied 
bands of BeTe. Note that the assignment given in the figure is not in 
all cases unambiguously derived from the spectra but is also partly based 
on the comparison of experimental and calculated spectra (see below). In 
addition to these bands there is a peak at $E_B=-5.5$ eV that remains 
unchanged in position as the photon energy is varied. For $h\nu > 15$ eV
two additional peaks are observed between binding energies of 0 and -2 eV (each
indicated with d) that are not assigned to direct transitions between bulk 
bands; they show some dispersion away from the energetic position of the 
valence band maximum; their origin will be discussed below.\\

For a detailed interpretation of the spectra in Fig.~2, the different peaks have
to be assigned to k-conserving transitions from occupied bands to unoccupied 
bands. The kinetic energy and the momentum of the electron in the vacuum are 
related according $E_{kin}=(\hbar^2/2m_{el})(k_\perp^2+k_\parallel^2)$, where 
$k_\parallel$ and $k_\perp$ are the components 
of the momentum of the electron in the vacuum, parallel and perpendicular to 
the surface, respectively and $m_{el}$ being the electron mass. The parallel 
component of the momentum is conserved (modulo a reciprocal surface 
lattice vector $\vec{G}_\parallel$) when the electron crosses the surface, 
and can therefore be directly determined from the momentum in the vacuum. 
To determine the perpendicular momentum inside the crystal, we must know the 
final state bands along the (100) direction. The simplest approximation to 
the final state band is the free electron approximation. The potential step at 
the surface is usually taken into account by introducing the inner potential, 
$V_o$, as the zero for the energy scale (see eg. Refs. 4-8). For normal 
emission, the parallel component vanishes ($k_\parallel=0$) and one 
obtains the electron momentum perpendicular to the surface as:\\

\begin{equation}
k_{final} = \frac{\sqrt{2m_{el}}}{\hbar}\cdot\sqrt{E_{kin}+V_o} + G_B,
\end{equation}

with $G_B$ being a reciprocal bulk lattice vector. While the assumption of a 
free electron parabola as the final state with the inner potential $V_o$ as an 
adjustable parameter is widely used in the analysis of photoemission spectra, 
there is no general justification for using this simplified approach and its 
applicability has to be verified for each system separately. Its success 
depends on the kind of a solid and the energy range of the final states.\\

In a more sophisticated approach one can analyze the measurements with the aid 
of a first principles calculation of the band structure of BeTe. The need of 
this kind of approach is strongly suggested by the dispersion of the unoccupied 
bands in Fig.~1, which is far from having a free-electron-like shape. Before 
the analysis was done, the unoccupied bands from the LDA calculation have been 
shifted by 1.1 eV to higher energies in order to fit the experimental absolute 
energy gap of 2.7 eV. However, even with the assumed first-principles final 
states a determination of the initial states is not straightforward. Due to 
a complicated shape of the dispersion of the conduction-bands  in BeTe there 
are in general several bulk excited states which could be final states 
for a given photon energy.\\

In order to obtain an unambiguous interpretation of the experimental spectra, 
we use an even more stringent criterion, namely the direct comparison of 
experimental and calculated photoemission spectra for increasing photon 
energies. For this purpose we calculate the electron-hole excitation rate for 
BeTe bulk, which determines to a large extent the emitted photocurrent 
(Eq.~(1)).  Fig.~3 shows the calculated bulk contribution to the 
photocurrent (solid lines) along with the experimental spectra. The calculated 
curves were rescaled in intensity. Overall, the agreement between calculation 
and experiment is very good, in spite of the applied approximations. 
The energetic positions of most of the peaks are very close to maxima in the 
experimental curves and the overall shape of the spectra is well reproduced. 
By evaluating Eq.~(1) and analyzing the theoretical spectra in more detail we 
are able to trace back the origin of particular photoemission peaks in Fig.~3. 
The dotted line stems from the transitions from the lower 
spin-split occupied bands, the dashed line represents only the transitions from
the upper bands (light and heavy holes together). The solid line is sum of 
both.\\

The experimentally determined valence band structure of BeTe between
$\Gamma$ and $X$ points is presented in Fig.~4 along with the calculated bands 
of Fig.~1. Only those unoccupied bands are depicted, which due to their 
symmetry properties could be observed in normal emission (see Ref.\ 15). They 
are shifted upward by 1.1 eV to correct for the LDA error (see above). The 
experimental data points have been obtained from the photoemission spectra 
in Figs.~2 and 3 using 
the calculated final states by determining the $k_\perp$ value from the 
intersection of the unoccupied band with a horizontal line (dotted lines in
Fig.~4) for the corresponding kinetic energy. In the following we will discuss 
the spectra in detail and will give the assignment of the various peaks in 
Fig.~3 that leads to the experimental band structure in Fig.~4. The four rather
strongly dispersing structures that have been assigned to direct, band-to-band 
transitions, have been labeled with full and empty circles and squares. In 
addition there are some non dispersive features. Note that the same symbols are
used to indicate the peak position in Figs.~2 and 3 and the binding energies
in Fig.~4.\\

Both the experimental and the calculated curves in Fig.~3 are rather 
sensitive to the photon energy. Up to photon energies of 11.5 eV, the 
most intense structure in both experimental (full circle) and 
calculated spectra (dashed curve) is close to zero binding energy. 
It corresponds to transitions from the uppermost occupied states (heavy
and light hole bands) around the $\Gamma$-point. 
At around 12 eV, a peak around $E_B=-1$ eV becomes the dominant 
structure in the experimental (full square) as well as in the 
calculated curves. From the calculation (dotted line) we conclude that 
its origin are transitions from the lower spin-split bands close to 
the $\Gamma$-point. 
Starting around $h\nu=13$ eV, an additional intense peak (at 
about $-2$ eV) can be clearly identified in the calculation (dashed 
line) as well as in the experiment (open circle). 
This peak is also present at lower photon energies (albeit with much 
lower intensity) and exhibits strong dispersion as the photon energy is
increased, from $E_B=-0.5$ eV at $h\nu=11$ eV to close to $-3$ 
eV at the highest photon energies. 
It is assigned to transitions from the upper occupied band in an 
intermediate $k_\perp$ region between the $\Gamma$ and $X$ point to 
final state band 4 in Fig.~4. 
Beginning at $h\nu=14$ eV the peak at $E_B=0$ eV (full circles) 
exhibits a shoulder that develops into a clearly dispersing peak, 
which has the largest intensity in the calculation (dashed line). It is
attributed to transitions from the upper occupied bands along 
$\Gamma X$ to final state band 5.
A very similar behavior is seen for the band at $-1$ eV (full squares, 
dotted line):  Starting at $h\nu=15$ eV a shoulder is observed 
at the high binding energy side that disperses to higher binding energy
as the photon energy is increased. It is assigned to transitions from 
the lower occupied band to final state band 5 in the first half of the 
Brillouin zone. \\

As a next step we will discuss the origin of the two only weakly dispersing 
peaks (indicated with d) between binding energies of $0$ and $-2$ eV, in the 
experimental as well as the calculated spectra for higher photon energies. 
Based on the calculation we assign them to transitions from the upper and the 
lower occupied band close to $\Gamma$ to different final states: Due to their 
short lifetime the final states are significantly broadened
(from about $0.25$ eV to $3$ eV for excitation energies between $10$ and $40$
eV\cite{c17}), 
and therefore transitions close to $\Gamma$ are observed throughout the photon 
energy range studied owing to the existence of various final state bands 
between $8$ and $16$ eV (bands 3-6) and the high density of occupied as well 
as unoccupied states in the neighborhood of this point. It is worth noting 
that the presence of almost non-dispersive features around the valence band 
maximum is rather commonly observed in photoemission experiments for other 
II-VI, III-V and elemental semiconductors\cite{c7,c19,c20}. A first and most 
often assumed interpretation is that these peaks are due to 
surface states. This is a plausible interpretation and the fact that our bulk 
calculation for larger photon energies ($>16$ eV) does not reproduce the 
experimental spectra as well as for lower energies seems to approve this point 
of view. First of all, there is no direct emission from
bulk states in this energy range. Secondly, for higher photon energies the 
inelastic mean free path of excited photoelectrons is smaller and therefore, 
the experiment could be more sensitive to the surface region. Moreover, in the 
case of CdTe (100) surface, calculations of the surface electronic structure 
predict a surface resonance in this energy range\cite{c21}.
On the other hand, a rather universal character of this feature, showing up in 
many systems with different geometric and electronic structures of the 
surfaces, as well as in the results of our strictly bulk calculation, seem to 
indicate the many-body (broadening of final
states) origin of this feature. Our assumed rigid shift of the empty bands is 
certainly a worse approximation for higher bands than lower\cite{c22}, which 
together with the neglect of the scattering character of final states could be 
responsible for the poorer agreement between our calculation and experiment for
higher photon energies. It is possible that both effects play here a role and 
on the basis of our experiment and calculation we cannot
make a definitive statement about their relative weight. The existence of 
surface states for BeTe(100) will be a subject of a future study\cite{c23}.\\

At this point we want to come back to the rather strong sensitivity of both 
the experimental and calculated spectra to photon energy below $h\nu=14$
eV and show that in this photon energy region the assignment of experimental 
peaks to particular transitions has to be performed with great caution: 
This becomes evident, if we analyze in more detail the behavior of the 
dispersing structure denoted with full circles in Fig.~3, which corresponds
to transitions from the upper occupied bands. For $h\nu=10$ eV the final
states for this structure are provided by the unoccupied band 3 
in Fig.~4. Increasing the photon energy to $11$ eV, the final states are given 
by the empty band 4 and the full circles in Fig.~3 show a dispersion towards 
the zero binding energy which is reached at $h\nu=11.5$ eV. Strictly 
speaking, the final state energy for the full circle structure at 
$h\nu=11.5$ eV is a small fraction of electronvolt {\it above} the 
calculated (and shifted) LDA band 4 in Fig.~4. However, due to the lifetime 
broadening (see above) of this band this transition is still very pronounced, 
as is clearly seen in the calculated spectrum of Fig.~3. For $h\nu=12$ 
eV, the final state energy lies between final state bands 4 and 5 around 
the $\Gamma$ point in Fig.~4. Nevertheless, the lifetime broadening allows to 
observe a peak with a binding energy somewhat below the VBM and clearly smaller
intensity; it is of the same origin as the weakly dispersing peaks at higher 
photon energies (see above) and is therefore also indicated with "d" in Fig.~3.
The second time the VBM is reached at $h\nu=12.5$ eV in both 
experiment and calculation. For this and higher photon energies the intensity 
of the peak (indicated with full circles) increases, which is related to the 
fact that now the strong transitions to the empty band 5 occur. The behavior 
of the peak at $-1$ eV binding energy (full squares) can be understood along 
the same lines.\\

We next discuss the peak that is marked with open squares and which is first 
observed at $E_B=-2.8$ eV for $h\nu=12$ eV and disperses to higher 
binding energy as the photon energy is increased. For the highest photon 
energies it merges into the non-dispersing peak at $E_B=-5.5$ eV. This 
dispersing feature is assigned to 
transitions from the lower occupied band in the second half of the Brillouin 
zone, i.e. closer to the $X$-point into unoccupied band 3. While the binding 
energies in calculation and experiment are in nearly perfect agreement for this
structure, the experimental intensity is larger than expected from the 
calculation. For bulk initial and final states with no spin-orbit interaction 
and our nominal experimental setup ($\vec{E}$ vector of the photon field in the
surface plane) these transitions are symmetry forbidden. By considering the 
spin-orbit interaction, the selection rules are relaxed and emission from the 
lower band is allowed, the transition amplitude being however very small. One 
reason for the pronounced, experimentally observed intensity of this peak could
be a non-perfect experimental setup, i.e. due to the integration over an angular
range of 3 degrees and a possible misalignment of the sample that should
however be smaller than 5 degrees. To simulate this situation, the calculation
in Fig.~3 has been performed with the light incoming to the sample with an 
angle of  5 degrees with respect to the surface normal. Another reason of the 
enhancement of the intensity of this peak could be the larger inelastic mean 
free path $\lambda_f$ of final states corresponding to this transition as 
compared to peaks of lower binding energy. Since the inelastic mean free path 
of excited electrons shows a characteristic minimum at about $50$ eV and the 
final states in our experiment correspond to the rather steeply descending part
of the $\lambda_f$ versus excitation energy curve, this many-body effect
also turns out to be important here. We have included in our calculation this 
effect as well, but nevertheless, the experimental intensities remain still
larger. It cannot be excluded that the final states relevant for these 
transitions couple stronger to free-electron states outside the sample and could
be easier detected than our bulk calculation can predict.\\

Finally, we discuss the strong dispersionless structure at $E_B=-5.5$ eV that 
is observed in the experimental spectra. A similar, non-dispersive structure 
has also been observed in other II-VI semiconductors and is usually 
interpreted as originating from indirect transitions from 
a high-density-of-state region around the $X$ point\cite{c24}. This 
interpretation is very plausible, because the binding energy of $-5.5$ eV 
agrees very well with the band structure calculation. In our calculation of 
photoemission spectra from the bulk band structure, no indirect transitions are
allowed and therefore a dispersionless feature is not observed. One 
should, however, mention that a contribution from a surface state, which has 
been proposed for various other compound semiconductors in that energy range,
cannot be completely ruled out.\\
 
\section{Summary and Conclusions}

We have investigated the band structure of BeTe along $\Gamma X$ using 
angle-resolved UV-photoelectron spectroscopy and theoretically determined 
photoemission spectra based on a LDA band structure calculation. The agreement 
between the experimental and calculated spectra is very good and allows 
a detailed assignment of the various photoemission peaks. The experimentally 
determined band structure of the occupied bands that has been 
obtained using the calculated unoccupied band structure is in excellent 
agreement with the calculated occupied band structure as is evident from 
Fig.~4. This is a strong indication for the fact that the LDA band structure
of both occupied bands and unoccupied bands when 
corrected by a uniform shift, is a good 
description of the electronic structure of BeTe in the excitation energy
range relevant for our experiment.\\

We want to mention that an interpretation of the experimental photoemission 
spectra of BeTe(100) based on the assumption of free electron final states 
does not lead to meaningful results\cite{c25},
at least in the photon energy range applied in the present study. This is 
immediately obvious from the inspection of the unoccupied bands in Figs.~1 
and 4. None of the bands can be approximated by 
a free electron parabola and therefore the interpretation based on this 
assumption necessarily has to fail. However, we would like to stress here 
an even more important aspect of this problem: with a proper choice of the 
inner potential $V_o$, which plays the role
of an empirical parameter for the assumed free-electron final states, one
could obtain a plausible picture of the occupied band structure. The case
of BeTe shows, however, that the assignment of the origin of
photoemission peaks in such an approach would be for many photoemission
features different from the correct assignment obtained within our 
first-principles scheme.\\

The assignment of the various peaks in the experimental spectra is considerably
simplified by comparison to the calculated ones. This is particularly 
true for peaks that have only small intensity. An other 
important point that we would like to stress and that has been often neglected 
in the past is the broadening of the final states that leads to non-dispersing 
or only weakly dispersing states due to transitions close to the $\Gamma$-point
in the calculated as well as the experimental spectra. For BeTe these effects 
explain the weakly dispersing peaks close to the valence band maximum 
for high photon energies, although other explanations 
cannot be excluded.\\
 
\section{Acknowledgments}
This work was supported by the BMBF through grant 05 622 WWB and by the 
DFG through SFB 410. We want to thank H. Koschel for his assistance and 
Prof. D. Menzel and Dr. W. Widdra for the cooperation and support when 
using their photoelectron spectrometer. The calculations have been performed 
at the Leibnitz Rechenzentrum in Munich.

\par

\newpage
\begin{figure}
\caption{%
Calculated band structure of BeTe(100) along the $L$- and $X$ direction in the 
Brillouin zone. The energy scale has been adjusted to zero at the valence-band
maximum at the $\Gamma$-point.}
\end{figure}

\begin{figure}
\caption{%
Overview of angle resolved valence band photoemission spectra of BeTe(100) 
collected at normal emission for increasing photon energies. The positions of 
four vertical band to band transitions are indicated with different symbols, 
two peaks that are attributed to the high density of states close to the 
$\Gamma$-point are indicated with d; for details see text.}
\end{figure}

\begin{figure}
\caption{%
Comparison of the experimentally determined BeTe(100) valence photoemission 
spectra of Fig.\ 2 (data points) with the calculated electron-hole excitation 
rate (solid line). The dashed line describes the contribution from the upper 
occupied bands (light and heavy holes together) and the dotted line is obtained
by the transitions from the lower (spin-split) occupied band. The positions 
of four vertical band to band transitions are indicated by different symbols, 
two peaks that are attributed to the high density of states close to the 
$\Gamma$-point are indicated with d; for details see text.}
\end{figure}

\begin{figure}
\caption{%
Comparison of the experimentally determined valence band dispersion $E(k_\perp)$
along the (100) direction of BeTe with the LDA band structure calculations. The
experimental data are obtained using the experimental spectra in Figs.\ 2 and 3
and the calculated unoccupied bands of Fig.\ 1 shifted to higher energies by 
$1.1$ eV (see text); only those final state bands are shown that can actually 
be observed in normal emission\cite{c16}. The unoccupied bands are labeled as 
bands 1-6 with increasing energy. Note that the same symbols are used 
in Fig.\ 2-4. As examples three transitions are indicated with vertical 
arrows.}
\end{figure}


\begin{thebibliography}{99}
\bibitem{c16} When the spin-orbit interaction is included all bands along the
$\Gamma X$ direction have the $\Delta_5$ symmetry and have no purely symmetric
or antisymmetric character, because the spin-orbit perturbation mixes both
kind of states from the unperturbed system. However, a symmetry/antisymmetry
character of most states is to a large extent preserved. Therefore, as a 
criterion for the selection of the photoemission-final states we have taken
the average value of the modulus square of each state along the symmetry axis:
when this value for a given state showed smaller than a chosen threshold, 
we have neglected such a state in the calculation of the photoemission
intensities.
\end{thebibliography}
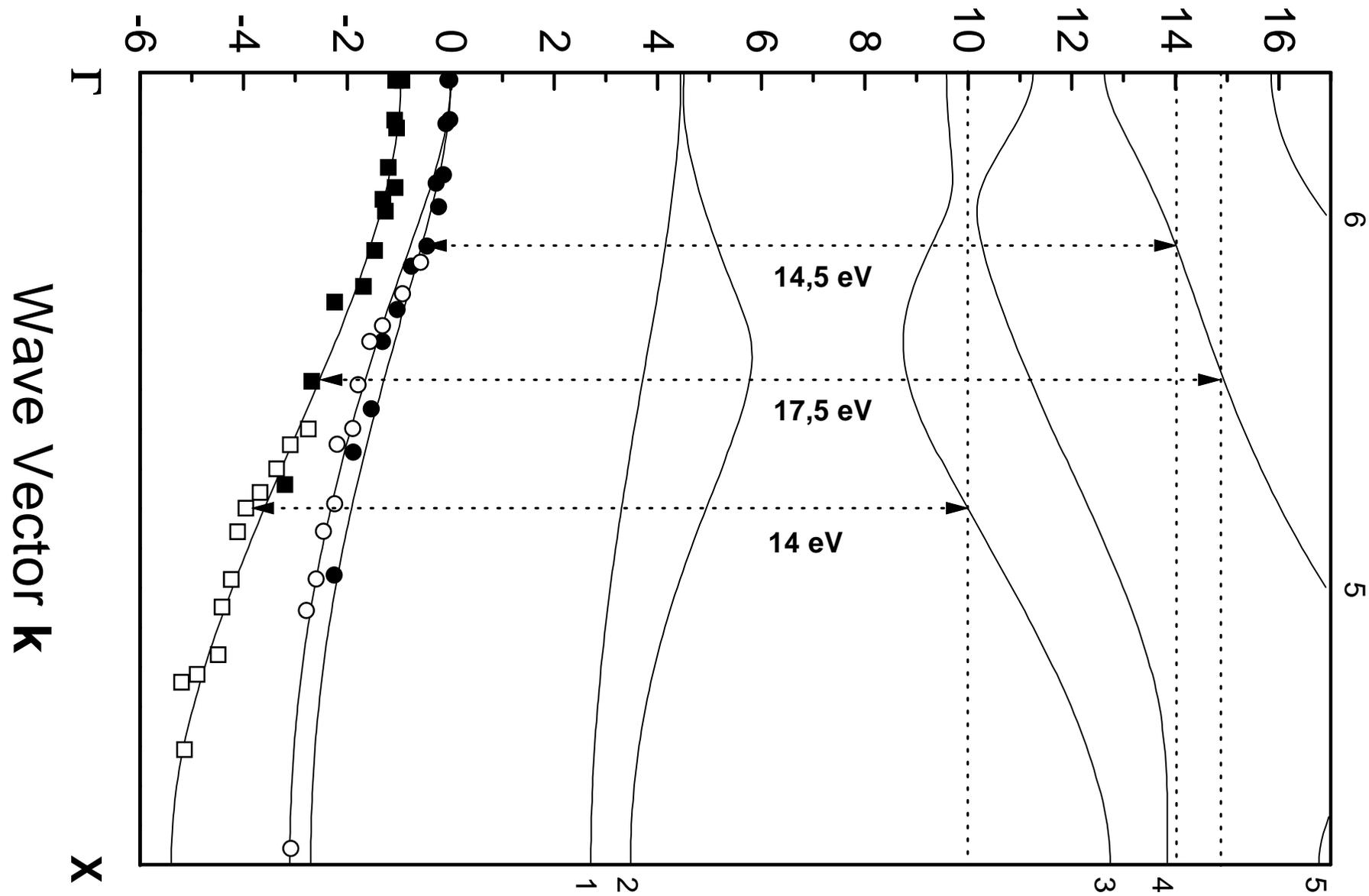
\end{document}